# A Divisive Hierarchical Clustering-Based Method for Indexing Image Information

Najva Izadpanah

Department of Computer Engineering,
Islamic Azad University, Qazvin Branch, Qazvin, Iran

*ABSTRACT*

*In most practical applications of image retrieval, high-dimensional feature vectors are required, but current multi-dimensional indexing structures lose their efficiency with growth of dimensions. Our goal is to propose a divisive hierarchical clustering-based multi-dimensional indexing structure which is efficient in high-dimensional feature spaces. A projection pursuit method has been used for finding a component of the data, which data's projections onto it maximizes the approximation of negentropy for preparing essential information in order to partitioning of the data space. Various tests and experimental results on high-dimensional datasets indicate the performance of proposed method in comparison with others.*

## 1. INTRODUCTION

Due to the advances in hardware technology and increase in the production of digital images in various applications, it is necessary to develop efficient techniques for storing and retrieval of such images [1, 2, 3].

One of the current techniques for image retrieval is content-based image retrieval. In content-based image retrieval approach, visual features such as colour feature, texture feature, shape feature and local features are automatically extracted from the image objects and organized as feature vectors. Then at search phase, after selecting the query image by user, retrieval engine retrieves the most similar images to the query image by performing similarity comparison between query feature vector and all the feature vectors in database [1, 4].

One of the discussed issues in content-based image retrieval research area is to use efficient methods to accelerate search operation in image retrieval databases. For this purpose, multi-dimensional indexing structures have been used for organization of multi-dimensional image features and then searching image database has been done via multi-dimensional indexing structure for having appropriate speed in retrieval process [5, 6, 7].

Up to now many multi-dimensional indexing structures have been proposed [8, 9]. But due to large volume of images in image databases and high dimensionality of image feature vectors, proposing an efficient multi-dimensional indexing structure capable of managing these large feature vectors is very arduous and important. Increase of the dimensionality strongly aggravates the quality of the multi-dimensional indexing structures. Usually these structures exhaust their possibilities until dimensions around 15[6].

However, it is shown that in high-dimensional data spaces, multi-dimensional indexing structures tend to perform worse than the sequential scanning of the database. This fact is due to the curse of dimensionality phenomena [7]. Hence new approaches have been recently proposed for indexing of multi-dimensional objects especially for high-dimensional spaces. In [9] a classification of high-dimensional indexing structures has been proposed.

Many clustering algorithms are proposed by now, but only few research studies are conducted to propose clustering methods in the purpose of applying them to indexing of image information [10].

Researchers have classified clustering methods in four total classes of partitioning algorithms, hierarchical algorithms, grid-based algorithms and density-based algorithms [11].

Hierarchical algorithms has developed in two classes of hierarchical divisive algorithms including PDDP[12] and hierarchical Agglomerative algorithms including BIRCH[13] and CURE[14] for partitioning the data space in a hierarchical structure.

The goal of this research study is to propose a divisive hierarchical clustering-based multi-dimensional indexing structure in order to manage high-dimensional image feature vectors which also prevents overlapping in its structure.

The rest of the paper is organized as follows: In section 2, we describe circumstance of the operation of retrieval system according to our proposed indexing structure. In section 3, proposed multi-dimensional indexing structure (NO-NGP-tree1) is described. In section 4, implementation and experiments on our proposed method are explained. And Section 5 includes conclusions and feature works.

## 2. CIRCUMSTANCE OF THE OPERATION OF RETRIEVAL SYSTEM ACCORDING TO OUR PROPOSED INDEXING STRUCTURE

Like image retrieval systems architecture proposed in [16], image retrieval based on our proposed indexing structure consists of four main components including Feature extraction, Datasets, Retrieval and Multi-dimensional indexing scheme. *Figure* .1 shows the image retrieval basic architecture and our proposed indexing scheme is shown with green color on it.

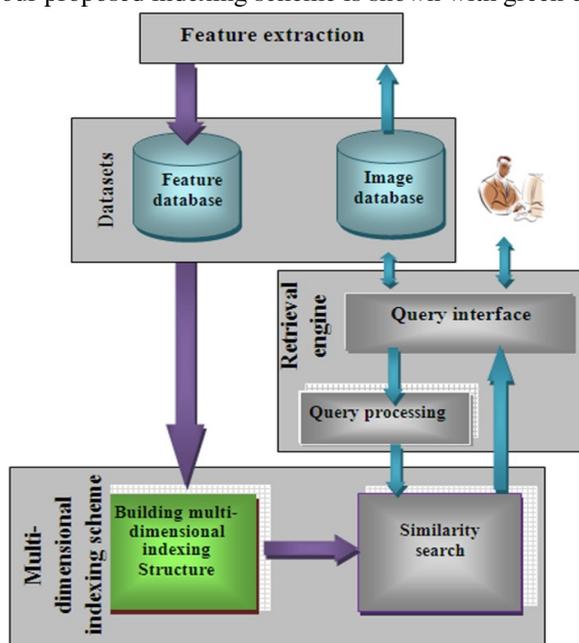

Figure.1    Image retrieval basic architecture and our proposed indexing scheme place in it

---

[1] No Overlapping-Non Gaussian Projection based-tree

Feature vectors are extracted from each image in the image database and saved in the feature database. When users ask their query via Query interface, image query is converted to a multi-dimensional feature vector via Query Processing. In Multi-dimensional indexing scheme, in an offline phase(Building multi-dimensional indexing structure phase), every feature vectors in feature database are organized in a hierarchical divisive clustering-based structure that we name it NO-NGP-tree, and then in similarity search phase, query feature vector is taken and via a recursive similarity search algorithm the NO-NGP-tree is searched and k-nearest neighbors to the query feature vectors send to the Query interface for representing to user.

In this research, result of the Building multi-dimensional indexing structure phase, as it is shown in Figure. 2, is an un-balanced binary tree that in highest level or root consists of all multi-dimensional feature vectors.

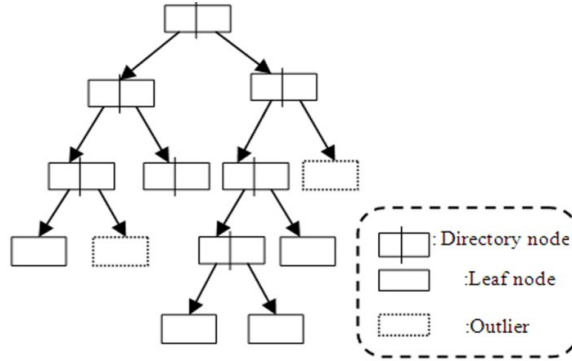

Figure.2    NO-NGP-tree structure

In NO-NGP-tree, nodes with two child are directory nodes and nodes with no child if have less data points than *Minpts* are outlier and otherwise are leave nodes.

In similarity search phase we used the nearest-neighbor search proposed in [17].

## 3. PROPOSED MULTI-DIMENSIONAL INDEXING STRUCTURE NO-NGP-TREE

In image retrieval system based on our proposed indexing structure, each image is represented as a multi-dimensional feature vector and all feature vectors are saved in the feature database. Therefore feature database could be considered as matrix $M_0$ (n, m) that consists of feature vectors in its rows. Where n is dimension of feature vectors and m is numbers of feature vectors in feature database.

In order to build the NO-NGP-tree, $M_0$ has partitioned recursively via an iterative algorithm. In each iteration a not complete binary structure has constructed. The iterative procedure continues until k leaf nodes are created. Finally a binary structure which is shown in Figure. 2 is built.

Pseudo code of NO-NGP-tree is shown in Figure. 3.

```
NO-NGP-tree (M_i, K, Minpts)
Begin
i =1
1. {I_i}= Pre-partitioning(M_i)
2. {(MR_i)_s, (ML_i)_s} =Partitioning (M_i ,I_i)
3. { (BR_i)_s,(BL_i)_s ,(MR_i)_s, (ML_i )_s }= Bounding ((MR_i)_s,
(ML_i)_s, (a_i)_s)
4. {TS_i}= Build tree structure((BR_i)_s,(BL_i)_s ,(MR_i)_s, (ML_i )_s)
5. Check if  LNo_i< k then go to 1 else go to 6, i=i+1
6. Return( NO-NGP-tree)
End
```

Figure. 3    NO-NGP-tree Pseudo code

In Figure. 4 the process of building proposed multi-dimensional indexing structure is shown. Process of building proposed multi-dimensional indexing structure consists of four main phase: Pre-partitioning, Partitioning, bounding and Build tree structure. This procedure iteratively continues until $k$ leaf nodes are created.

Since data nodes of each leaves could be considered as a matrix, in each iteration, all leaf matrixes of tree in iteration ith that is defined by set $M_i$ is taken. $M_i$ has been defined as in (1).

(1) $M_i = \{M_{i-1} - (M_{i-1})_s\} \cup (MR_{i-1})_s \cup (ML_{i-1})_s - outlier_{i-1}$   Where, $M_{i-1}$ is the set of leave matrixes at i-1th iteration, $(MR_{i-1})_s$ and $(ML_{i-1})_s$ are right sub-cluster and left sub-cluster that is selected at i-1th iteration, and $outlier_{i-1}$ is an outlier at i-1th iteration.

Also $(I_i)_j$ has been defined as in(2).

(2) $(I_i)_j = \{F_j, (a_i)_j, IDX1, IDX2, CP1, CP2\}$

Where, $F_j$ is a vector consists of projections of jth leaf on to the $(a_i)_j$, $(a_i)_j$ is the meaningful non-gaussian component of jth leaf in $M_i$, IDX1, IDX2 are projection sub-clusters, and CP1,CP2 are centroids of projection sub-clusters which all will explain later in this paper.

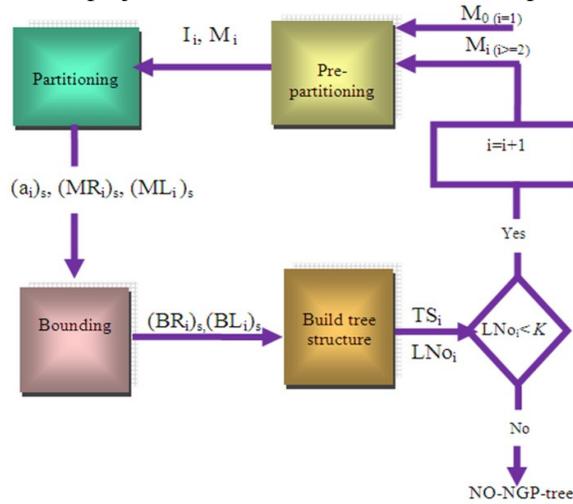

Figure. 4 Process of building NO-NGP-tree

In the first iteration (i=1, i is the counter for numbers of iterations), $M_0$ is given as input. In Pre-partitioning phase, required information for partition phase ($I_i$) is prepared. Then in partitioning phase, next cluster for farther partitioning will be selected and partitioned into the right sub-cluster($(MR_i)_s$) and left sub-cluster($(ML_i)_s$).

Partitioning is done based on projection of datanodes of selected cluster on the meaningful non-Gaussian component obtained from pre-partitioning phase $((a_i)_s)$. Then $(a_i)_s$ will be sent to the bounding phase. In the bounding phase, right and left sub-clusters will be bounded with minimum bounding rectangles that directed according to the $(a_i)_s$, therefore there is no overlap between them. Results of bounding phase are $(BR_i)_s,(BL_j)_s$ that are bounding information about left and rightsub- clusters of selected cluster in ith iteration.

Then in tree structure building phase, two right and left sub-clusters will be verified and if number of datanodes in each of them is less than *Minpts* ,it will be marked as outlier ,otherwise it will be marked as leaf node. In both cases $(BR_i)_s,(BL_j)_s$ will be saved. Then $TS_i$ that specifies the binary tree in ith iteration in addition to $LNO_i$ will be resulted in this phase. It should be said that $LNO_i$ is the number of leaf nodes in ith iteration.

## 3.1 Pre-Partitioning

In Pre-partitioning phase, required information for partition phase ($I_i$) is prepared. In this phase, the $M_i$ consists of leaf matrixes in i-1 iteration is taken and then for each member of $M_i$ that are leaf nodes, in three phase, $(I_i)_j$ will be obtained. In this research, $(I_i)_j$ for j=1 to $LNO_i$ are defined as $I_i$, where j is the counter for numbers of leaves in $M_i$, and $(M_i)_j$ is the jth leaf in $M_i$.

Then $I_i$ and $M_i$ will be sent to partitioning phase.

As the pre-partitioning pseudo code shows in Figure. 5, pre-partitioning consists of following three phase that iterate for j=1to $LNO_i$:

- Find meaningful non-gaussian component: In this phase the meaningful nono-gussian component will be extracted from $(M_i)_j$ where this component that is specified with $(a_i)_j$ directed according to a direction that shows the clustered structure of data distribution.
- Projection: In this phase, data nodes In $(M_i)_j$ will be projected on to the $(a_i)_j$.
- 2-mean clustering: In this phase, 2-mean clustering will be applied on projections, then two centroids (IDX1, IDX2) and two clusters (CP1, CP2) will be obtained that we specified these two sub-clusters with projection sub-clusters.

```
Pre-partitioning (Mi)
Begin
1.For j=1 to LNOi  Do:
Begin
1.1{(ai)j} = Find  meaningful non-gaussian component
((Mi)j)
1.2{ Fj }=Projection((ai)j,(Mi)j):
1.3{CP1, Cp2, IDX1, IDX2}= 2-mean clustering (Fj)
End
2.Return(Ii)
End
```

Figure. 5 Pre-partitioning Pseudo code

### 3.1.1 Find meaningful non-gaussian component

For building proposed multi-dimensional indexing structure, in each iteration selected leaf matrix will be partitioned in to two sub-clusters at partitioning phase. For this purpose, data nodes of selected leaf have been projected on to one component extracted from that leaf, and then at partitioning phase have been decomposed in to two sub-clusters from a data node among projected data nodes.

Idea of projecting data on to a component from data is not a new idea and it has already been used in hierarchical divisive algorithms such as PDDP. In PDDP, the data set is divided into two sub clusters. In such a way that, it projects all the data points on to first principal component. After the algorithm has split the initial dataset into two sub clusters, it is going to recurse this procedure for one of the two sub clusters. This partitioning strategy creates a binary tree whose leaves are the final clusters of the data set.

In this research, in order to propose a hierarchical divisive clustering method, a Projection Pursuit(PP) method has been used for finding a meaningful non-gaussian component of the data which projections of data onto maximizes the approximation of negentropy, therefore data projections on this component could contain desirable information for splitting data into two sub-clusters[12]. In fact, in PDDP, $PCA^2$ [18, 19] linear transformation is done and dataset is projected on first principal component.

Desirability of the components(directions) with maximum non-gaussian attribute than principal components(directions) is shown in Figure. 6.

In Figure. 6, first principal component which is shown with green color presents the clustering structure of the data distribution, but non-guassian component that is shown with red color presents the separation between clusters then it shows the clustering structure of data distribution.

---

[2]Principal Component Analysis

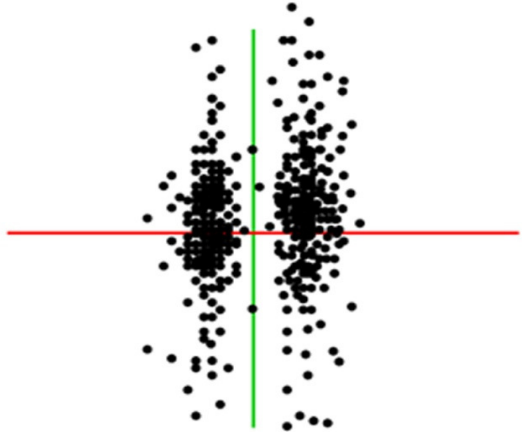

Figure. 6  Desirability of non-gaussian components than principal components.

The main idea of Projection Pursuit methods is to extract meaningful component from dataset with local optimization of a desire Objective function (Projection index) [20]. Past studies indicate that Gaussian distribution shows less meaning at data distribution and desire components (directions) have maximum distance from Gaussian directions.

Local optimization of an Objective function is also done in ICA[3] which is one of the Projection Pursuit methods.

In [21] using transformation of neural network learning rules to fixed point iteration, FastICA algorithm has been proposed. FastICA is originally proposed for finding independent components but in fact, it is a Projection Pursuit method that we can use to find meaningful non-gaussian component.

FastICA algorithm finds each component individually. Therefore we can use it for finding only one component. Objective function in this algorithm could be each objective function which is used by researchers in process of finding projection pursuits. In this research, we use one version of FastICA algorithm that approximation of negentropy is used in as objective function.

Negentropy is one of the most important measures of non-*gaussianity*.

Gaussian variables have most entropy among variables with same variance. Therefore, negentropy could be considered as one of the non- *gaussianity measures*. Data distribution with more gaussianity and therefore with more entropy is more non-structured and more unpredictable. Negentropy J is defined as in (3) [20].

(3) $\quad J(y) = H(y_{gauss}) - H(y)$

Where, $y_{gauss}$ is Gaussian variable that has same variance with variance of $y$, and H is entropy.

Since it is difficult to calculate negentropy, approximation of negentropy practically has been used as Objective function in [20]. This approximation defined in (4).

(4) $\quad J(y) \cong [E\{G(y)\} - E\{G'(V)\}]^2$

Where, E is expectation function, V is a Gaussian variable with zero mean, and G is defined in (5).

(5) $\quad G(u) = \tanh(cu)$

Where $1 \leq c \leq 2$ is a constant that is usually set as 1.

In this research, FastICA with first principal component as initial weight vector is used to extract a meaningful non-gaussian component and the first component found by the algorithm is used for projecting of data onto, because projection of data onto it maximizes the approximation of negentropy and therefore since this component is directed according to the direction of data with most non-gaussianity attribute, projections of data onto it could include interesting information for partitioning the dataset into two sub-clusters.

---

[3] Independent Component Analysis

### 3.1.2 Projection

In projection phase, data matrix $(M_i)_j$ ($M_0$ in first iteration) will be projected on to $(a_i)_j$. projection on $(a_i)_j$ is characterizes in (6).

(6) $$F_j(x_b) = (a_i)_j^T (x_b)$$

Where, $x_b$ is bth feature data point(image feature vector) in $(M_i)_j$, and $F_j(x_b)$ is a vector whose each component is a scalar that shows the projection value of $x_b$ ($b$=1 to nodesize) on to $(a_i)_j$, nodesize is the number of data points in $(M_i)_j$.

Data Projections on to the meaningful non-gaussian component could consists of following information:

- Locations with less density are appreciating locations for partitioning of data points of related cluster from them.

- Locations with more density are close to the sub-cluster centroids of related cluster.

In Figure. 7 an example of not balanced and not separated clusters has been shown. In Figure. 7, first principal component is shown by green color and the meaningful non-gaussian component is shown by red color. We can see that the red one is aligned to the direction with more irregularity. Then projection of data on the red one could show the locations with more and less density.

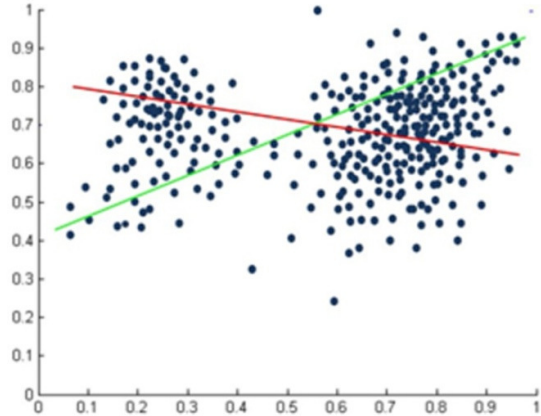

Figure.7 an example of not balanced and not separated clusters

### 3.1.3 Clustering of data projections

In this phase, a clustering algorithm such as k-means [22] will be applied on the data projections $F_j$. As the result two projection sub-clusters (IDX1, IDX2) and their centroids (CP1, CP2) will be obtained.

Data projections onto $(a_i)_j$ could show the locations with more and less density, But how we can find these locations. In this research, in order to guess these location, k-means clustering with k=2 has been applied on projections. So two obtained centroids are considered as locations with more density, and the mean of these two centroids is considered as the location with less density.

In data distribution, with balanced and separated natural clusters, means of two centroids of projection sub-clusters is probably near the real centroid of the data but in data distribution with not balanced and not separated natural clusters , the mean of two projection sub-clusters centroids is shown the location with less density that centroid of the data.

Figure. 8 shows centroids of the dataset shown in Figure. 7, and the means of projection sub-clusters centroid. As it is shown in Figure. *8*, location of means of the projection sub-clusters centroids that located on the meaningful non-gaussian component has less density than the location of the real data centroid.

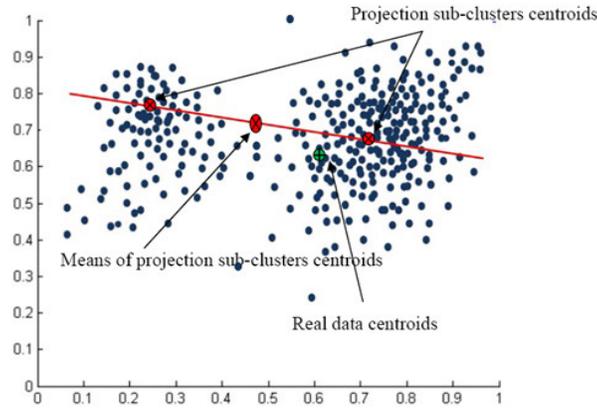

Figure. 8 comparing location of means of projection sub-clusters centroids with the location of real data centroid

## 3.2 Partitioning

In partitioning phase, $I_i$ and $M_i$ is taken. $M_i$ includes leaves of tree structure constructed until ith iteration. Figure. 9 shows partitioning pseudo code.

```
Partitioning (M_i ,I_i):
Begin
1.(M_i)_s = Cluster selection (M_i,I_i)
2.{(MR_i)_s, (ML_i)_s }=Split( (M_i)_s , I_i)
3.Returns((MR_i)_s, (ML_i)_s)
End
```

Figure. 9  Pre-partitioning Pseudo code

As the partitioning pseudo code shows in Figure. 9, partitioning consists of following two phases:
- Cluster selection: In this phase, one of the leaf node clusters in ith iteration will be selected for further partitioning. In this research the node with more clustered structure will be selected.
- Split: In this phase, the selected cluster will be split into two sub-clusters.

### 3.2.1 Cluster selection

In this phase, the cluster which should be split into two sub-clusters is selected from clusters related to the leaf nodes. For this purpose Mi and Ii has been taken and then according to this information one of the clusters related to leaf node will be selected.

In this research, information obtained from projection of data of each cluster on to the $(a_i)_j$ has been used and a cluster selection measure is proposed.

Cluster selection is one of the main challenges of divisive clustering methods. In some past studies (for exapmle see [12]), the cluster with the most scatter value has been selected for further partitioning. Scatter value is defined as in (7).

(7) $$Scattvalue = \frac{1}{N}\sum_{i=1}^{N}\|x_i - w\|^2$$

Where, *Scatvalue* is scatter value, N is the numbers of data points in related cluster, and w is the centroid of the cluster.

But as it is shown in [23], scatter value is not a good measure for measuring the clustering structure of a cluster.

Proposed cluster selection measure is defined in (8).

(8) $$selvalue = \left|CP1 - CP2\right| \Big/ \max_{c=1,2}(diameter(IDX_c))$$

Where, CP1 and CP2 are projection sub-clusters centroids obtained from the considered leaf node, and diameter (IDXc) is the diameter of cth(c=1,2) projection sub-cluster diameter that is defined in (9).

(9) $$diameter(IDX) = \max(|F_p|) - \min(|F_p|)$$

Where, IDX is the considered projection sub-cluster, and $F_p$ is pth data projection.

Longer distance between centroids of two projection sub-cluster with less cluster diameter indicated the more clustering structure of the data. In other words, the cluster with more value of this measure has more clustering structure.

In Figure. 10, a simple example with its dataponits, their projections on to the meaningful non-gaussian component , and centroids of projection sub-clusters is shown.

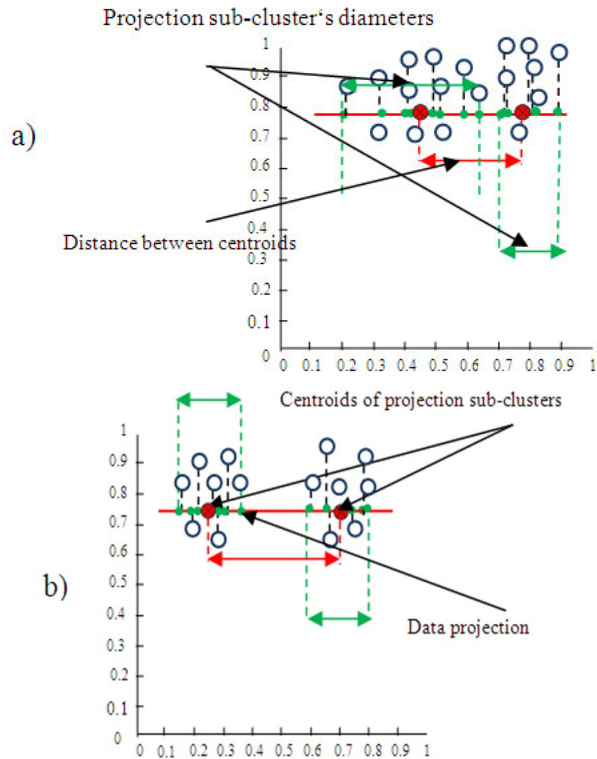

Figure. 10 an example of data projections on to the meaningful non-gaussian component in two-dimensional space. a) cluster with unstructured distribution, selvaue=1.5. b) cluster with structured distribution, selvalue=2.8.

Figure. 10.a shows a cluster with unstructured distributation, which has the selvalue equal to 1.5, and Fig.10.b presents a cluster with structured distribution and the selvalue equal to 2.8.

**3.2.2 Split**

In this phase, the matrix related to the selected leaf in ith iteration($(M_i)_s$) and its meaningful non-gaussian component ($(a_i)_s$), as well as the vector consists of projection data of selected leaf ($F_s$) are taken. Then it is split into two right and left sub-clusters from the means of centroids of projection sub-clusters by the hyper-plane orthogonal to the $(a_i)_s$. $(a_i)_s$ is also saving for right and left sub-clusters.

In section 3.1.3, we have discussed about why it is better to split the cluster by the hyper plane passing from the means of centroids of projection sub-clusters.

Figure. 11-a), shows the separating hyper-plane passing from real (origin) centroids of the data distribution was shown in Figure. 7. As it is demonstrated in Figure. 11 because the data distribution consists not balanced and not separated clusters, real centroids of the data is not a good choice as location that separating hyper plane passing from , but mean of the centroids of projection sub-clusters is more suitable. Figure. 11-b) shows the separating hyper-plane passing from mean of the centroids of projection sub-clusters for the data distribution shown in Figure. 7.

As it is shown in Figure. 11, by hyper-plane orthogonal to the meaningful non-gaussian component passing from mean of the centroids of projection sub-clusters, the data is partitioned into two tight sub-clusters, and finally after bounding it is resulted to the tighter minimum bounding rectangles.

As the result of splitting by hyper-plane orthogonal to the meaningful non-gaussian component passing from mean of the centroids, data points in selected cluster that their projections are in the right side of the hyper-plane will be assigned to the right sub-cluster that attributed by $(MR_i)_s$, and data points that their projections are in the left side of the hyper-plane will be assigned to the left sub-cluster that attributed by $(ML_i)_s$. In fact, for each $x_b$, if result of (10) is positive, then $x_b$ will be assigned to right sub-cluster and, else if it is not positive then $x_b$ will be assigned to the left sub-cluster.

(10) $\quad F_s(x_b) - (a_i)_s^T c_{mean}$

Where, $x_b$ is $b$th data point, $F_s(x_b)$ is the projection of $b$th data point, and *Cmean* is the mean of centroids of projection sub-clusters.

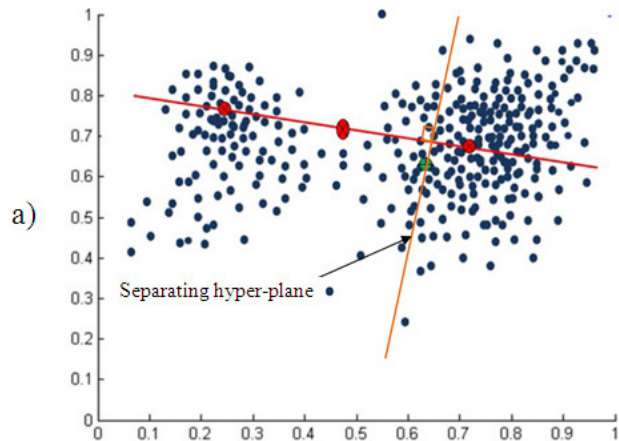

a)

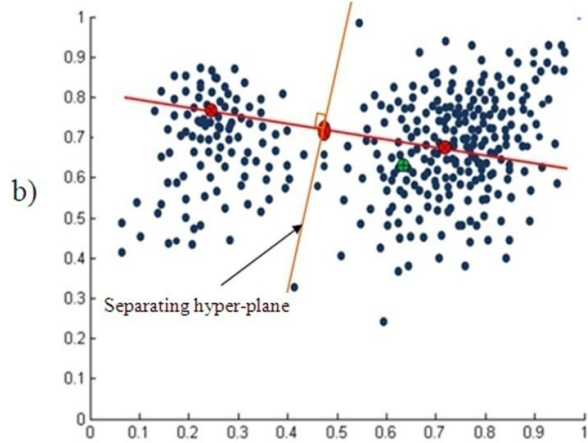

Figure. 11 Split by separating hyper-plane. a) hyper-plane orthogonal to the meaningful non-gaussian component passing from mean of the centroids of projection sub-clusters. b) hyper-plane orthogonal to the meaningful non-gaussian component passing from the real data centroid.

### 3.3 Bounding

As the process of building NO-NGP-tree is shows in Figure. 4, in each iteration, after partitioning phase, right and left sub-clusters obtained from partitioning phase and also $(a_i)_s$ will be sent to the bounding phase, and in bounding phase, right and left sub-clusters will be bounded with minimum bounding rectangles which are directed according to $(a_i)_s$.

Bounding pseudo code is shown in Figure. 12.

```
Bounding ((MR_i)_s, (ML_i)_s, (a_i)_s)
Begin
1. {(a_i)_s, ,(NR_i)_s, (NL_i)_s }=Change-reference mark
   ((MR_i)_s, (ML_i)_s, (a_i)_s)
2. {(BR_i)_s, (BL_i)_s } =Minimum bounding rectangle
   construction ( (NR_i)_s, (NL_i)_s)
3. Returns((BR_i)_s,(BL_i)_s ,(MR_i)_s, (ML_i)_s)
End
```

*Figure*r. *12* **Bounding pseudo code**

Bounding phase consists of two following phases:
- Change-reference mark: in this phase, reference mark is changed according to the $(a_i)_s$ direction.
- Minimum bounding rectangle construction: in this phase, minimum bounding rectangles constructed in the new reference mark.

#### 3.3.1 Change-reference mark

As it has been mentioned before, to prevent from bounding overlapping, both right and left sub-clusters are bounded with minimum bounding rectangles which are directed according to $(a_i)_s$ therefore bounding rectangles of two nearby nodes do not have any overlap.

Changing reference mark and directed bounding rectangles according to the new reference mark in order to prevent from overlapping covering nodes area is not a new idea. In [17], changing reference mark is done in NOHIS-tree for preventing from nodes overlapping.

In this phase, right and left sub-clusters in new reference mark obtained using (11) and (12).

(11) $(NR_i)_s = (MR_i)_s - 2V^T.V.(MR_i)_s$

23

(12) $(NL_i)_s = (ML_i)_s - 2V^T.V.(ML_i)_s$

Where, $(NR_i)_s$, $(NL_i)_s$ are matrixes related to the right and left sub-clusters in new reference mark, and $V = \frac{((a_i)_s - e_1)}{\|(a_i)_s - e_1\|}$ where $e_1 = (1,0,....,0)$.

### 3.3.2 Minimum bounding rectangle construction

In this phase right and left sub-clusters in new reference mark are taken and bounded with minimum bounding rectangles (MBRs). Then information related to the right and left MBRs which are defined with $(BR_i)_s$ and $(BL_i)_s$ will be saved. In this research, for construction of MBRS the method that it is used in [17] has been applied.

It is shown in Figure. 13 that how changing reference mark and constructing minimum bounding rectangles according to the new reference mark leads to minimum bounding rectangles of nearby nodes with no overlap between them.

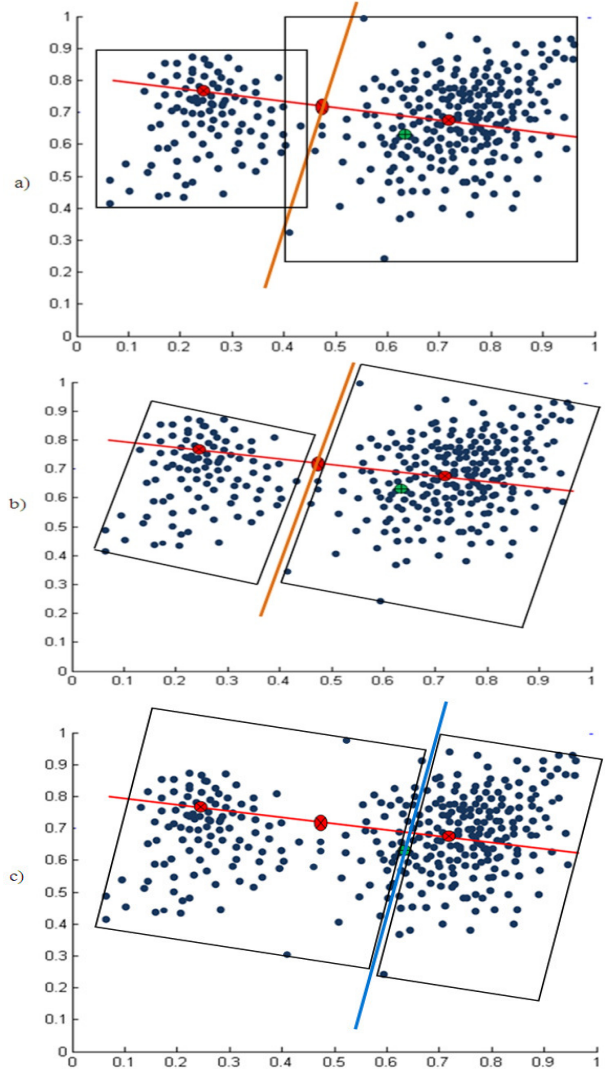

Figure. 13 MBRs construction. a) MBRs construction according to real reference mark. b) MBRs construction according to the new reference mark. c) data splitting by the heyper-plane passing through real data centroid and MBRs constructing according to the new reference mark

As we can see in Figure. 13-b) construction of MBRs according to the new reference mark prevents overlapping between nodes. In Figure. 13-a) and b) data is split by the hyper-plane orthogonal to the meaningful non-gaussian passing through mean of centroids of projection sub-clusters. But in Figure. 13- c) data is split by the hyper-plane orthogonal to the meaningful non-gaussian passing through real data centroid, and MBRs are constructed according to the new reference mark.

As it is shown in Figure. 13- c) that there is no overlapping between bounding areas of the nodes, but as splitting is done by the hyper-plane passing through the real data centroid, it is lead to not completely separated subclusters and therefore the total volume of the bounding rectangles in Figure. 13-c ) is larger than total volume of bounding rectangles in Figure. 13-b).

### 3.4 Building tree structure

In this phase, in each iteration of constructing the proposed index structure, right and left sub-clusters and also information related to the right and left MBRs are added to the tree structure $TS_i$. Building tree structure pseudo code is shown in Figure. 14.

As we can see in Building tree structure pseudo code, if the number of data nodes in each sub-cluster was less than *Minpts*, it will be marked as outlier, otherwise it will be marked as leaf node.

```
Build tree structure ((BR_i)_s,(BL_i)_s, (MR_i)_s, (ML_i)_s)
Begin
1. If number of rows of ((MR_i)_s / (ML_i)_s) < Minpts
   Then let ((MR_i)_s / (ML_i)_s) be outlier
   Else let it be leaf
2. Add (MR_i)_s, (ML_i)_s to the binary tree structure and
   save (BR_i)_s, (BL_i)_s, LNO_i = LNO_i+1
3. Returns(TS_i, LNO_i)
End
```

Figure. 14. Building tree structure pseudo code.

## 4. IMPLEMENTATION AND EXPERIMENTS

We have reported implementation and experiments results of our proposed method and in this section.

### 4.1 Implementation

In this section, we have covered details about implementation such as evaluation measures, test method, datasets and implementation environment.

### 4.1.1 Evaluation measures

In this research, Recall and Response time are considered as evaluation measures of our proposed method.

Quality of final evaluated clusters is evaluated with Recall measure. Recall is considered as ratio of the numbers of data points related to the query points that are retrieved to the total numbers of related data points to the query point.

Recall is defined in (13) [10, 24, 25].

(13) $$\text{Recall} = \frac{|\alpha \cap \beta|}{|\alpha|}$$

Where, $\alpha$ is set of related data points and $\beta$ is set of related retrieved data points.

Also in this research the retrieval speed via proposed indexing structure is evaluated with Response time measure. Response time is the time that is expended to search the database via indexing structure for finding k-nearest data points in the database to the query point. Response time is defined in (14) [10, 17, 24].

(14)
*Response time = the time that final results exit from similarity search phase-the time that query point enter to the similarity search phase*

Less values for response time shows better performance of indexing structure.

### 4.1.2 Test method

All tests in this research are done with Cross-validation method. For this purpose, each test is done 10 times and each time , 20 feature vectors from dataset are considered as test set, and for each of them all the dataset is scanned and 20-nearest neighbors for them will be found and inserted in the related data point set ( α ). Then indexing structure will be constructed with the rest of feature vectors in the dataset. Then we use test set for queries and mean results of 20 queries will be obtained and finally mean result for 10 tests will be reported [24, 23].

### 4.1.3 Datasets

To evaluate and test our proposed method, we have used a database including 1,193,647 feature vectors extracted from color images.

Feature vectors are local features based on points of interest derived from 4,996 color images which are used in previous studies for evaluating indexing structures performance [17].

In this research, we have used four databases of different dimensions including 50,000 feature vectors from described dataset.

First database includes 50,000 feature vectors of dimension 25. Second database includes 50,000 feature vectors of dimension 40. Third database includes 50,000 feature vectors of dimension 60 and fourth database includes 50,000 feature vectors of dimension 80.

### 4.1.4 Implementation environment

In this research, for implementing indexing methods in our experiments, all algorithms are implemented in MATLAB 7.8.0.347. Algorithms run on a PC with Intel CPU.4GHZ , 4GB of RAM.

## 4.2 Experiments

In this section, we have covered details about our experiments on our proposed indexing method.

### 4.2.1 Prerequiste parameters adjustments

In this section, we have done experiments for adjusting prerequisite parameters of our proposed indexing scheme.

As it is described in pseudo codes of section 3, we have considered two prerequisite parameters $k$ and *Minpts.* Parameter  $k$ is final numbers of clusters (final leafs and outliers). In fact parameter $k$ defines the numbers of leaf nodes plus outlier nodes after building of NO-NGP-tree. Also in our proposed scheme, minimum number that is considered for numbers of data nodes in each leaf node is defined with parameter *Minpts*. In our experiments, *Minpts* characterized as percent of average numbers of final clusters data members. For example 15 mean 15 percent of average numbers of final clusters data members.

Average numbers of final clusters data members are defined in (15).
(15)

*Average numbers of final clusters data members= total numbers of data nodes /k*

Where, *k* is final numbers of clusters.

For adjusting two parameter k and *Minpts*, some experiments are done and the average Response time is reported. These experiments that their results are shown in continue are done with different values for *Minpts* and fixed value for *k* in each for database of dimensions 25, 40, 60 and 80.

Results of average Response time obtained from experiments on NO-NGP-tree based on the second are shown in Tab. 1.

Table.1 Average Response time for different values of *Minpts* and *k* on databases with different dimensions

| 65 | 45 | 35 | 25 | 15 | 5 | *Minpts* |
|---|---|---|---|---|---|---|
| Average Response(s) time on different dimensional databases ، *k=200* | | | | | | |
| 0.146 | 0.125 | 0.112 | 0.104 | 0.125 | 0.146 | 25-d database |
| 0.427 | 0.386 | 0.375 | 0.352 | 0.368 | 0.462 | 40-d database |
| 0.613 | 0.593 | 0.558 | 0.535 | 0.578 | 0.663 | 60-d database |
| 0.732 | 0.703 | 0.679 | 0.664 | 0.684 | 0.77 | 80-d database |
| Average Response(s) time on different dimensional database ، *k=600* | | | | | | |
| 0.074 | 0.068 | 0.063 | 0.062 | 0.069 | 0.084 | 25-d database |
| 0.316 | 0.257 | 0.23 | 0.228 | 0.243 | 0.36 | 40-d database |
| 0.438 | 0.41 | 0.372 | 0.345 | 0.36 | 0.475 | 60-d database |
| 0.607 | 0.561 | 0.521 | 0.513 | 0.54 | 0.632 | 80-d database |
| Average Response(s) time on different dimensional databases ، *k=800* | | | | | | |
| 0.124 | 0.11 | 0.103 | 0.098 | 0.109 | 0.173 | 25-d database |
| 0.401 | 0.35 | 0.321 | 0.301 | 0.343 | 0.447 | 40-d database |
| 0.46 | 0.451 | 0.387 | 0.36 | 0.406 | 0.524 | 60-d database |
| 0.617 | 0.58 | 0.532 | 0.52 | 0.567 | 0.65 | 80-d database |
| Average Response(s) time on different dimensional databases ، *k=1000* | | | | | | |
| 0.201 | 0.155 | 0.148 | 0.14 | 0.145 | 0.246 | 25-d database |
| 0.563 | 0.526 | 0.49 | 0.453 | 0.518 | 0.61 | 40-d database |
| 0.763 | 0.724 | 0.683 | 0.652 | 0.705 | 0.729 | 60-d database |
| 0.97 | 0.946 | 0.9 | 0.873 | 0.925 | 1.026 | 80-d database |

Results in Tab.1 shows that with growing of dimensions of data, average Response time is increased for different values of *Minpts*. Also it is considered from results in Tab.1 that with variation of values for *Minpts*, different results have been obtained. Therefore, for analyzing the results more accurately, in Figure. 15 with waiver of number of dimensions, histogram of average Response times on 25-d database for different values of *k* is shown.

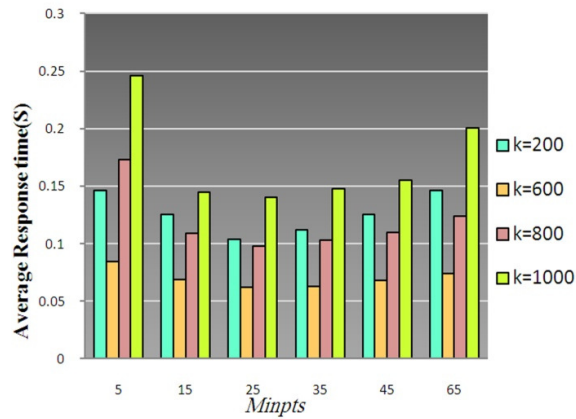

*Figure. 15* Histogram of average Response times on 25-d database for different values of *k*

Figure. 15 demonstrates that with adjustments done on prerequisite parameters, when *Minpts* is 25 and k is 600, best results are obtained. Also it is observed that when *Minpts* is adjusted on greater values such as 65 or less value such as 5, average response time is increased. Also when Signal & Image Processing : An International Journal (SIPIJ) Vol.6, No.1, February 2015*Minpts* is adjusted on 15 to 25, better results have obtained. In fact if a minimum value considered for number of leaf node data members (*Minpts*) in each iteration is very large, since cluster selection measure used in our scheme tends to select smaller clusters, small clusters will be split continuously. This leads to increase in height of the unbalanced indexing structure and therefore to decrease in speed of search via proposed indexing structure. Also adjusting *Minpts* on very small values leads to not fine detection of real outliers and just clusters with very small numbers of data nodes are marked as outliers, so that not incorrect partitioning is done and then it leads to decrease in the quality of final clusters as well as increase in Response time.

Also it is shown in Figure. 15 that Response time for different values of *k* and fixed value of *Minpts* is different. This is because if the value of *k* is near to the real number of natural clusters in dataset, quality of the final clusters is better, and since it leads to search in fewer clusters in our scheme it leads to decrease of Response time through our proposed indexing structure.

**4.2.2 Experiments for evaluating quality of final searched clusters**

In this section, results of experiments for evaluating quality of final searched clusters with Recall measure in compare to following schemes are reported:

NGP-tree: this structure was obtained through elimination of changing reference mark phase from the process of constructing the proposed indexing structure NO-NGP-tree.

PDDP-tree: this structure was obtained by adding MBRs clustering method PDDP.

NOHIS-tree: a clustering based indexing method [17].

It is observable that in four considered indexing methods, in similarity search phase, we have used similarity search algorithm that was proposed in [17].

Experiments on NO-NGP-tree and NGP-tree in this section are done with adjusting *Minpts* on 25 and different values for *k*. The purpose of running the experiments with different values of parameter *k* is to shows the effect of adjusting this parameter in a comparison of similar indexing schemes.

Figure. 16 shows average Recall with respect to number of searched final clusters on 25-d and 80-d databases where *k* is 600,800 and 1000, and *Minpts* is 25.

Via NO-NGP-tree, searching the database after searching average 14 final clusters is stopped. But via NOHIS-tree after searching average 20 final clusters is stopped.

As it is demonstrated in Figure. 16, with increasing the number of searched clusters, average Recall is increased, and after searching about 14 clusters via NO-NGP-tree Recall turns to 1.
Also it is demonstrated in Figure. 16 that for PDDP-tree and NGP-tree in whose structures there are some overlaps between nodes, average Recall is increased slightly, and it is stopped for NGP-tree after searching about 257 clusters and for PDDP-tree after 273 clusters. Slower increase in the average Recall for structures with some overlaps is actually because of not accurate calculation of MINDIST (distance between MBRs and query vector [17]). Also if MBRs are tighter, the calculation of MINDIST will be more accurate. Figure. 16 indicates that NO-NGP-tree has better results than NOHIS-tree because of optimum partitioning during building indexing structure which leads to tighter MBRs.

As we can see in Figure. 16, average Recall on 80-d database for four methods increases slower than 25-d database. Also with increasing the value of prerequisite parameter *k*, average Recall for NO-NGP-tree and NGP-tree has less increasing, because when *k* has more distance from real numbers of natural clusters in dataset distribution, quality of final clusters will be decreased. This decrease in the performance shows itself on 80-d database in such a way that after searching 12 clusters, average Recall for NOHIS-tree is higher than average Recall for NO-NGP-tree. Reason of this problem is decrease in performance of the algorithm used in pre-partitioning phase for finding a meaningful non-gaussian component from data on high dimensional databases.

28

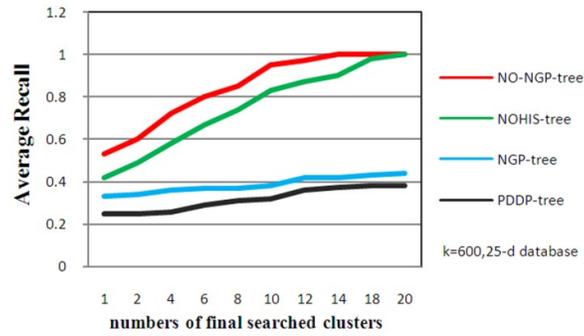

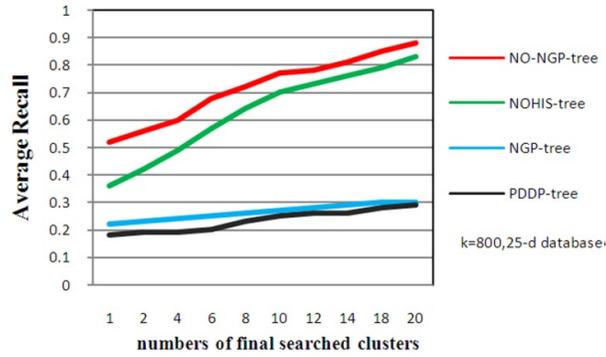

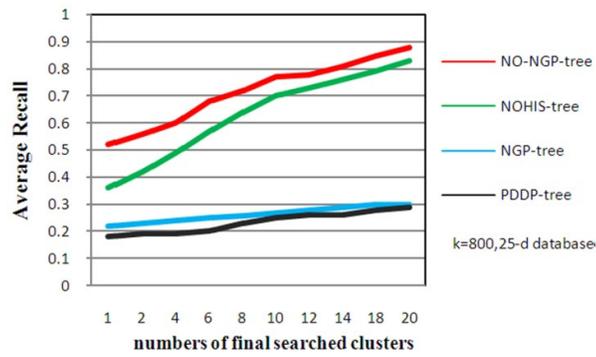

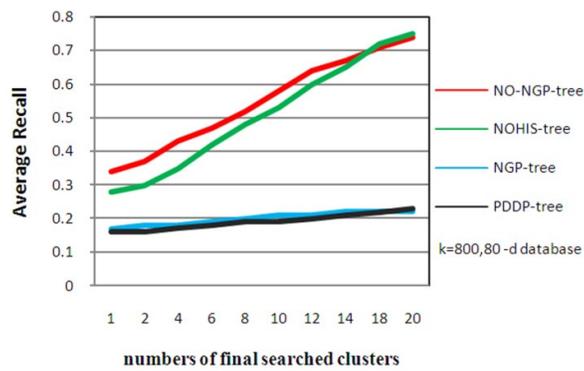

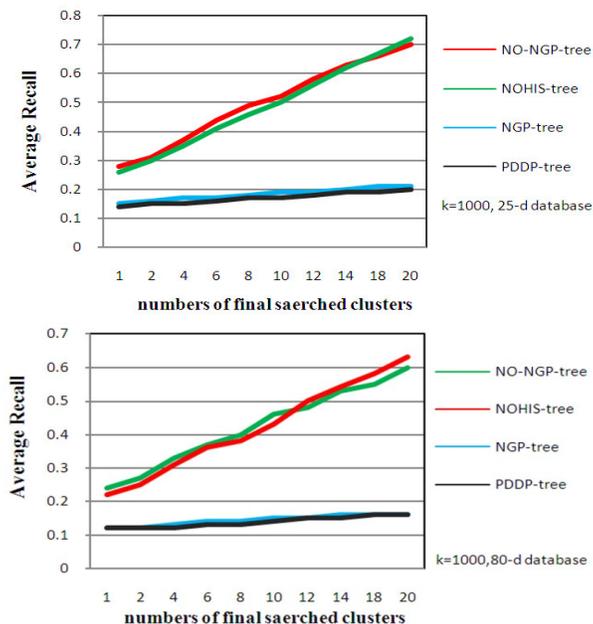

Figure. 16 average Recall respect to numbers of final searched clusters on 25-d and 80-d databases where k is 600, 800 and 1000, and Minpts is 25

## 4.3 Experiments for evaluating search speed via proposed indexing scheme

In this section, results obtained from our experiments for evaluating search speed via our proposed indexing structure with Response time evaluation measure in compare to NOHIS-tree, NGP-tree, PDDP-tree and sequential-scan is reported.

Experiments in this section are done by adjusting prerequisite parameters for NO-NGP-tree and NGP-tree on best case obtained from adjusting parameters section. Therefore in all experiments of this section, *Minpts* is adjusted on 25 and *k* is adjusted on 600. Also, in similarity search phase, we have used similarity search algorithm that was proposed in [17].

Figure. 17 shows average Response time for the comparison of NO-NGP-tree, NOHIS-tree, NGP-tree and PDDP-tree on 25-d, 40-d, 60-d and 80-d databases.

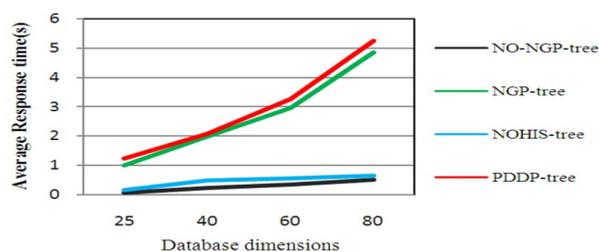

Figure. 17 average Response time for comparing NO-NGP-tree, NOHIS-tree, NGP-tree and PDDP-tree on 25-d,40-d, 60-d and 80-d databases

As it is demonstrated in Figure. 17, our proposed indexing scheme NO-NGP-tree that is shown in black color is has the less response time than NOHIS-tree, NGP-tree and PDDP-tree for different dimensions. Also it is demonstrated that Response time is growing with increase in dimensions. Because in these indexing schemes, increase in dimensions of database leads to increase in the volumes of minimum bounding rectangles (MBRs) of nodes. Therefore, average distance of query vector from MBRs will increase, and then it leads to decrease in accuracy of calculating MINDIST distance. This leads to more navigation of tree structure and also more final clusters that has been searched during search process then it leads to growing Response time.

Also Figure. 17 indicates that NGP-tree and PDDP-tree in whose structures, overlapping between nodes is not guaranteed, have more Response time than NO-NGP-tree and NOHIS-tree. Reason of this problem is the increase in number of navigations of tree structure because of these overlaps.

In this research, we claimed that reason of decrease in Response time via proposed NO-NGP-tree in compare to NOHIS-tree is optimum partitioning and tighter sub-clusters, which leads to decrease in the total volumes of MBRs in NO-NGP-tree structure.

Figure. 18 shows average Response time for comparison of NO-NGP-tree and Sequential-scan on 25-d, 40-d, 60-d and 80-d databases.

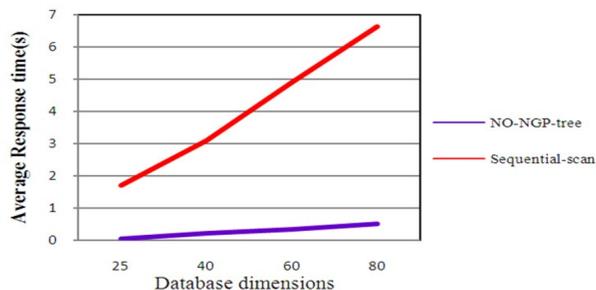

Figure. 18 average Response time for comparing NO-NGP-tree and Sequential-scan on 25-d, 40-d, 60-d and 80-d databases

We compared our indexing scheme with sequential-scan because most conventional indexing schemes lose their efficiency in high-dimensional spaces so that sequential scanning of dataset shows better results [6, 7].

As it is demonstrated in Figure. 18, NO-NGP-tree improves response time with large distance than sequential-scan.

## 5. RESULTS AND FUTURE WORKS

In most applications of image retrieval, images are described with high-dimensional feature vectors but research studies on indexing structures have shown that these structures lose their efficiency with increase in dimensions.

In this research we have proposed an indexing structure for managing high-dimension image feature vectors using pattern of hierarchical divisive clustering methods. In section 4, prerequisite parameters have adjusted on different values and our proposed scheme is evaluated and results are reported. Also we have run some experiments for evaluating final clusters quality which can also affect on search efficiency. Finally by adjusting prerequisite parameters on best case we have run some other experiments for evaluating speed of search in the database via our proposed indexing structure.

Some more investigations in this field which might be helpful in improvement of our indexing schemes in feature include:

1-Using other objective functions instead of approximation of negentropy in process of finding a meaningful non-gaussian component and compare their results.

2-Our indexing structure has an unbalanced binary structure. Therefore splitting the selected cluster to several sub-clusters instead of only two sub-clusters in each iteration might improve our indexing scheme performance.

3-Propose a method to estimate prerequisite parameter *k* value using selection model methods.

**REFERENCES**

[1]   Ritendra Datta, Dhiraj Joshi, Jiali, and James Z. Wang. Image Retrieval: Ideas, Influences, and Trends of the New Age, The Pennsylvania State University, ACM Transactions on Computing Surveys, 2008.
[2]   Shanmugapriya, N., and R. Nallusamy, A New Content based Image Retrieval System GMM and Relevance Feedback, Journal oF Computer Science,VOL.10,NO.2, 2014.
[3]   CHRISTIANBÖHM, STEFAN BERCHTOLD, DANIEL A. KEIM, searching in  high-dimensional spaces-Index structure for Improving the performance of Multimedia Databases,ACM Computing surveys,vol.33,No.3,322-373,2001.
[4]   M.S.LEW, N.SEBE, C.DJERABA, and R. JAIN, Content-based Multimedia Information Retrieval: State of the Art and Challenges, ACM Transactions on Multimedia Computing, Communications, and Applications, 2006.
[5]   V. Gaede and O. Günther, Multi-dimensional Access Methods, ACM Computing Surveys, Vol. 30, No. 2, 1998.
[6]   Krassimir Markov, Krassimira Ivanova, Ilia Mitov and Stefan Karastanev, ADVANCE OF THE ACCESS METHODS, International Journal of Information Technologies and Knowledge, Vol.2 , 2008.
[7]   Kaushik Chakrabarti, Sharad Mehrotra, The hybrid tree:an structure for high dimensional feature spaces,IEEE international confererance on data engineering,1999.
[8]   M-R.keyvanpour, N.izadpanah, Analytical Classification of Multimedia Index Structures by Using a Partitioning Method-Based Framework, The International Journal of Multimedia & Its Applications (IJMA) Vol.3, No.1, February 2011.
[9]   M-R.Keyvanpour,N.izadpanah, Classification and evaluation of High-dimensional Image Indexing Structures,The International journal of Machine learning and Copmuting,vol.2,No.3,2012.
[10]  C. Li and E. Chang and, H. Garcia-Molina and G. Wiederhold, Clustering for approximate similarity search in high-dimensional spaces, IEEE Transactions on Knowledge and Data Engineering, Vol.14,No.4,pp.792–808, 2002.
[11]  Hongli Xu, De Xu, and Enai Lin. An Applicable Hierarchical Clustering Algorithm for Content-Based Image Retrieval.Springer-Verlag Berlin Heidelberg, 2007.
[12]  DANIEL BOLEY, Principal Direction Divisive Partitioning.Data Mining and Knowledge Discovery 2, 325–344 (1998).
[13]  Tian Zhang, Raghu Ramakrishnan, and Miron Livny. BIRCH: An Efficient Data Clustering Method for VeryLarge Databases. In Proceedings of the 1996 ACM SIGMOD International Conference on Management ofData, Montreal, Canada, pp. 103-114, 1996.
[14]  Guha S, Rastogi R, Shim K. CURE: an efficient clustering algorithm for large databases, Proc. of the ACM SIGMOD Int'l Conf. on Management of Data. Seattle: ACM Press, pp. 73-84, 1998.
[15]  Aapo Hyvärinen and Erkki Oja, Independent Component Analysis: Algorithms and Applications. Neural Networks, Vol.13,No.(4-5),pp.411-430, 2000.
[16]  Y. Rui and T. S. Huang, Image Retrieval: current techniques,promising, Directions,and Open Issues. Journal of Visual communication and Image representation Vol.10, Issue. 1, pp.39-62,1999.
[17]  M. Taileb, S. Lamrous,and S. Touati, Non-overlapping Hierarchical Index Structure for similarity search International Journal of Computer Science, Vol. 3 ,No. 1, 2007.
[18]  J.E. Jackson. A User's Guide to Principal Components. New York: John Wiley and Sons,Inc, 1991.
[19]  I.T. Jolliffe. Principal Component Analysis.  Second edition, New York: Springer-Verlag New York, Inc, 2002.
[20]  Aapo Hyvärinen and Erkki Oja, Independent Component Analysis: Algorithms and Applications. Neural Networks, Vol.13,No.(4-5),pp.411-430, 2000.
[21]  Hyvärinen, A. and Oja, E., A fast fixed-point algorithm for independent component analysis. NeuralComputation, Vol .9,NO.7,pp:1483–1492, 1997.
[22]  J. MacQueen, Some Methods for Classification and Analysis of Multivariate Observations. In Proc. 5th BerkeleySymp. Mat. Statist, Prob., 1, pp. 281-297, 1967.
[23]  Sergio M. Savaresi, Daniel Boley, Sergio Bittanti, Giovanna Gazzaniga. Cluster Selection in Divisive Clustering Algorithms. In Proceedings of SDM',2002.
[24]  Dantong Yu and Aidong Zhang, ClusterTree: Integration of Cluster Representation and Nearest-NeighborSearch for Large Data Sets with High Dimensions, IEEE transaction on Knowledge and data Engineering, VOL. 15, NO. 3, MAY/JUNE 2003.
[25]  T.M. Mitchell, Machine Learning. McGraw-Hill, 1997.